\documentclass[12pt]{iopart}
\usepackage{graphicx}
\usepackage{amssymb}
\newtheorem{lemma}{\sc \bf Lemma}[section]
\newtheorem{propos}{\sc \bf Proposition}[section]
\newtheorem{theor}{\sc \bf Theorem}[section]
\newtheorem{corr}{\sc \bf Corollary}[section]
\newtheorem{remark}{\sc \bf Remark}[section]
\newtheorem{definition}{\sc \bf Definition}[section]

\begin{document}
\title[Self-similar formation of the Kolmogorov spectrum in the Leith model of turbulence]{Self-similar formation of the Kolmogorov spectrum  in the Leith model of turbulence}

\author{ S. V. Nazarenko$^1$ and
V. N. Grebenev$^2$}

\address{$^1$ Mathematical Institute, University of Warwick,
Coventry CV4~7AL, UK}
\address{$^2$ Institute of Computational Technologies SD RAS,
Lavrentiev avenue 6, Novosibirsk 630090, Russia}

\ead{ S.V.Nazarenko@warwick.ac.uk; vngrebenev@gmail.com}

\begin{abstract}
The last stage of
evolution toward the stationary Kolmogorov spectrum of hydrodynamic turbulence is studied using the Leith model~\cite{Leith} . This evolution is shown to manifest itself as a reflection wave in the wavenumber space propagating from the largest toward the smallest wavenumbers, and  is described by a self-similar solution of a new (third) kind.
This stage follows the previously studied stage of an initial explosive propagation of the spectral front from  the smallest to the largest wavenumbers reaching  arbitrarily  large wavenumbers in a finite time, and which was described by a self-similar solution of the second kind~\cite{Nazarenko,Nazarenko2,GNMS}. Nonstationary solutions corresponding to``warm cascades" characterised by a thermalised spectrum at large wavenumbers are also obtained.
\end{abstract}
\submitto{\JPA}
\maketitle

\section{Introduction}

Remarkably many fundamental properties of the hydrodynamic turbulence can be understood based on the
the simplest phenomenological model of Leith~\cite{Leith} in which the energy spectrum $E(k,t)$ obeys a nonlinear diffusion equation
\begin{equation}\label{I1}
\frac{\partial E}{\partial t} = \frac{1}{8}\frac{\partial}{\partial k}\left(k^{11/2}E^{1/2}\frac{\partial}{\partial k}(E/k^2)\right) - \nu k^2 E,
\end{equation}
where $t$ is time, $k$ is the absolute value of the wavenumber and $\nu$ is the kinematic viscosity coefficient. This is a special case of the singular nonlinear inhomogeneous diffusion equations, see e.g.~\cite{Vazquez}.

The Leith model is based on the assumption that the noninear interactions are local in the scale space, and it represents a minimal model that respects the scalings of more compicated turbulence closures. In particular, in the inertial range (when the viscosity term can be neglected) equation~(\ref{I1})  admits two fundamental stationary scaling solutions: the thermodynamic spectrum, $E(k) \sim k^2$, and the Kolmogorov spectrum, $E(k) \sim k^{-5/3}$. These scaling solutions are ``built into" the model, but they are not the only fundamental properties described by equation~(\ref{I1}), i.e. the Leith model is  essentially predictive and not merely descriptive.

An immediate prediction of the Leith model which was not put into it by the construction is the general inviscid
 steady state---a nonlinear combination of the thermodynamic and the Kolmogorov scalings~\cite{Nazarenko}:
\begin{equation}\label{01}
E_{P,T}(k) = ck^2(Pk^{-11/2} + T^{3/2})^{2/3},
\end{equation}
where $c = (24/11)^{2/3}$ and $P$ and $T$ are arbitrary constants corresponding to the energy flux through $k$ and a ``temperature". For $T = 0$, we recover the pure Kolmogorov cascade solution, whereas for $P = 0$---a pure thermodynamic spectrum.
Such solutions were called "warm cascade"
 in~\cite{Nazarenko} as they describe the so-called bottleneck phenomenon of spectrum stagnation near the cut-off  scale~\cite{Cichowlas:2005p1852} or a
crossover scale (e.g. classical-quantum crossover in superfluid turbulence~\cite{L'vov-Nazarenko-Rudenko}).

Another important prediction made with the help of the Leith model concerns
transient  solutions  arising from an initial spectrum compactly supported at low $k$ and preceding
formation of steady cascade. Provided that the initial conditions correspond to high Reynolds numbers, one can neglect viscosity in such transient evolution and use the inviscid Leith model:
\begin{equation}\label{invis}
\frac{\partial E}{\partial t} + \frac{\partial \epsilon}{\partial k} =0,
\quad \epsilon =- \frac{1}{8}k^{11/2}E^{1/2}\frac{\partial}{\partial k}(E/k^2),
\end{equation}
where $\epsilon$ is the energy flux.
Time-dependent solutions of this equation
 were investigated numerically in~\cite{Nazarenko,Nazarenko2} and analytically in ~\cite{GNMS}, and extensions to other turbulent systems (e.g. wave turbulence) were made in~\cite{Thalabard}. It was shown that the evolution becomes self-similar  just before breaking of energy conservation at some finite time $t=t_*$ at which the front of the spectrum reaches $k = \infty$. This is the so-called self-similarity of the second kind, using the Zeldovich-Raizer terminology~\cite{Zeldovich}. Remarkably, this regime does not exhibit the scaling inherited from the Kolmogorov spectrum.
Namely, the transient spectrum behind the propagating front was found to have a power-law asymptotics $E \sim k^{-x}$ with $x$ which is greater than the Kolmogorov exponent,  $x^*\approx 1.85>5/3$.
Previously, a similar behaviour of a transient spectrum exhibiting an anomalously steep power law was found numerically in MHD wave turbulence~\cite{galtier,nazar11}. A steeper transient spectrum was also found numerically for the EDQNM model of hydrodynamic turbulence
\cite{bos}, giving $x^*\approx 1.9$ which rather close to the exponent observed for the Leith model.
Moreover, a steep transient spectrum with  $x^*\approx 4$ was also found in direct numerical simulations (DNS) of the Euler equations for the ideal fluids \cite{brachet}.

Further, the Leith model was used to classify all possible types of behaviour in stationary turbulence with forcing and dissipation on the right or/and left boundaries of the $k$-range in~\cite{grebenev}. These include the Kolmogorov, thermodynamic and mixed solutions for low and high Reynolds numbers in the forward and inverse cascade settings which arise in the model~(\ref{I1}) with various types of the boundary conditions as $t \to \infty$.

On the other hand, there remain questions about the evolution for $t_*<t<\infty$.
Note that because we deal with a finite-capacity system, and because the evolution near $t=t_*$ is very fast at high $k$,
presence of the forcing and dissipation at the ends of the $k$-range is unimportant. Numerical simulations presented in~\cite{Nazarenko,Nazarenko2} reveal that the during this period of time there is a reflected wave propagating from large toward small $k$ into the power-law spectrum with steep exponent $x_*$ and leaving behind its front a shallower spectrum with a shallower power-law  spectrum whose exponent is very close to
Kolmogorov's $5/3$. Before that a similar scenario was observed in the numerical simulations of the wave-kinetic equation of weak MHD turbulence in~\cite{galtier}. However, such an evolution has not been yet explained theoretically. The main goal of the present paper is show that this final stage of the Kolmogorov spectrum formation  can be described by a self-similar solution of the {\em third kind} of the {\em inviscid} Leith equation~(\ref{invis}).

\section{On the classification
of self-similar solutions}

Zeldovich and Raizer ~\cite{Zeldovich} suggested the following classification. Self-similar solutions whose indices of self-similarity ($a$ and $b$ in our text below) are uniquely determined by a conservation law (i.e. effectively by the dimensional analysis) are of the first type.
Self-similar solutions for which the indices cannot be deduced for a conservation law or dimensional
analysis, and for determination of which one has to solve a nonlinear
eigenvalue problem are of the second type.

As we will see, the self-similar solutions considered in the present paper cannot fit in either of these two categories. Neither their can be fixed by a conservation law or dimensionally, nor they are determined by an eigenvalue problem solution. Instead, the self-similarity indices are fixed by a {\em prescribed } asymptotic behaviour at {\em one} of the ends of the self-similarity variable range. In the example of the reflection wave considered below this is the low-$k$ end, and the self-similarity indices are fixed by the exponent $x_*$ of the power-law spectrum ahead of the wave.

For the lack of an existing name, and following the Zeldovich-Raizer line of terminology, we will say that such self-similar solutions are of the
{\em third kind}. The first example of solution of this kind was
obtained, as far as we are aware,  in Ref.~\cite{Meerson}
for a dynamical cooling of a hot spherical air cavity.
 Note that the third-kind and the first-kind  solutions  share the property that they are defined for a formally unbounded time---unlike the second-kind solutions defined for a finite time range only.
(Of course physical relevance of such infinite-time self-similar solutions hold only for a finite time in most applications.) On the other hand,  the third-kind and the second-kind  solutions  share the property
that their indices are not determined by a conservation law or a dimensional analysis---unlike the  the first-kind  solutions.

\section{Self-similar solutions of the third kind}\label{S1}

Just before the blowup moment $t_*$,  the front of the spectrum reaches the dissipative wavenumber $k_\nu$ at which  viscosity $\nu $, no matter how small, is important.  However, at $k \ll k_\nu$ the evolution is still inviscid even for $t > t_*$: what happens at $k \sim k_\nu$ simply plays a role of an effective boundary condition for the low-$k$ dynamics.
After making this observation, we will study such a dynamics at
$t > t_*$ using equation~(\ref{invis}).

Equation~(\ref{invis}) admits  forms the following family of self-similar solutions:
\begin{equation}\label{1}
E = (t-t_*)^aF(\eta),\quad \eta = k/(t-t_*)^b,
\end{equation}
where $a$ and $b$ are constants called the self-similarity indices. They
satisfy the self-consistency condition,
\begin{equation}\label{2}
a = -2 - 3b,
\end{equation}
ensuring that equation for $F(\eta)$ is an ODE, namely
\begin{equation}\label{3}
-(3b + 2)F - b\eta\frac{dF}{d\eta} = \frac{1}{8}\frac{d}{d\eta}\left(\eta^{11/2}F^{1/2}\frac{d}{d\eta}(\eta^{-2}F)\right).
\end{equation}
Function $F(\eta)$ described by this equation describes evolution at $t >t_*$, but the boundary condition at  $\eta \to 0$ is determined
by the scaling at  $t=t_*$ forming during the pre-$t_*$ stage.
Namely, we look for a positive solution $F$ which behaves as $\eta^{-x^*}$ at $\eta \to 0$, where  $x^*$ is the exponent of the power-law forming at the
pre-$t_*$ stage, $t<t_*$, $t\to t_*$.  The numerical value found in~\cite{GNMS} is $x^* \approx 1.8509$. Further, because evolution at the low-$k$ part is much slower than at the high-$k$ part, the spectrum at $k \to 0$ may be considered stationary, $E=\hbox{const} \, k^{-x^*}$. This corresponds to ${x^*} = -a/b$. Together with condition (\ref{2}), this fixes the values of the of the self-similarity indices $a$ and $b$. Thus, the
self-similarity indices are fixed by the asymptotics at one of the ends, in this case at $\eta \to 0$, and this fits the definition of the self-similarity
of the third kind, as defined above. We have
\begin{equation}\label{ssindices}
a=-\frac{2x^*}{x^*-3}
 \quad \hbox{and} \quad
b=\frac{2}{x^*-3}
.
\end{equation}

To find $F(\eta)$, one must state one more boundary condition, e.g. at
$\eta \to \infty$. However, at this point we will not do that thereby leaving a one-parametric freedom in the shapes of $F(\eta)$. We will postpone the discussion about the relation of these shapes and conditions at the
high-$\eta$ end until later.

\begin{figure}[h]
 \center{ \includegraphics[width=12cm]{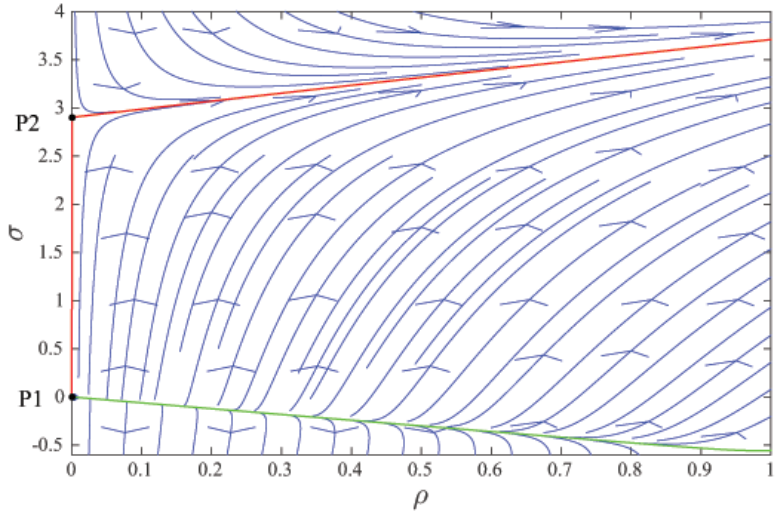} }
\caption{Phase portrait of the dynamical system. The slow (unstable) manifold of $P1$ (called $U1$) is shown in green. The heteroclinic orbit $H$ connecting $P1$ and $P2$, and the unstable manifold of $P2$ (called $U2$) are shown in red. }
\label{portrait}
\end{figure}

Written in terms of $x^*$ rather that $b$, equation~(\ref{3}) becomes
\begin{equation}\label{4}
-\frac{2}{x^*-3}\left(\eta\frac{dF}{d\eta} + x^* F\right) = \frac 1 8 \frac{d}{d\eta}\left(\eta^{11/2}F^{1/2}\frac{d}{d\eta}(\eta^{-2}F)\right).
\end{equation}
 Equation~(\ref{4})  can be transformed into an autonomous system
 by substitutions (c.f. \cite{Nazarenko}):
\begin{equation}\label{5}
F = \frac{8}{25}\eta^{-3}f^2, \quad \frac{dF}{d\eta} = \frac{24}{25}\eta^{-4}fg,
\end{equation}
where $f(s)$ and $g(s)$ are functions of $s = \ln\eta$. The resulting autonomous dynamical system is:
\begin{eqnarray}\label{6}
\frac{df}{ds} &=& \frac{3}{2}(f + g),\\
f\frac{dg}{ds} & =& \frac{1}{3}\left(5f^2 + 6fg - 9g^2 - \frac{10}{x^*-3}(3g + x^* f)\right). \nonumber
\end{eqnarray}
This system is singular at $f = 0$. By the change of variable
$$
\frac{d}{ds} = \frac{1}{f}\frac{d}{d\tau},\quad \rho(\tau) = f(s), \quad \sigma(\tau) = g(s),
$$
the system~(\ref{6}) is transformed to
\begin{eqnarray}\label{7}
\frac{d\rho}{d\tau} &=& \frac{3}{2}\rho(\rho + \sigma),\\
\frac{d\sigma}{d\tau} & =& \frac{1}{3}\left(5\rho^2 + 6\rho\sigma - 9\sigma^2 - \frac{10}{x^*-3}(3\sigma + x^* \rho)\right). \nonumber
\end{eqnarray}
Fixed points   of the system~(\ref{7}) in the  semi-plane $\rho \ge 0$   are
\begin{equation}\label{fixedpoints}
P1 =(\rho_1,\sigma_1) = (0,0) \quad \hbox{ and} \quad P2= (\rho_2,\sigma_2) = \left(0, \frac {10}3  (3-x^*) \right).
\end{equation}
A simple analysis reveals that $P1$ is an unstable saddle-node with its stable manifold along the $\sigma$-axis and its unstable (slow) manifold directed into the fourth quadrant with angle $-\arctan (x^*/3)$.
Fixed point $P2$ is a saddle with its unstable manifold along the $\sigma$-axis. The phase portrait of the dynamical system
is shown in Fig.~\ref{portrait}.

  At $\tau \to -\infty$ ($\eta \to 0$) we have $F \sim \eta^{-x^*}$ and $dF/d\eta \sim -x^*\eta^{-x^* - 1}$. It follows that $\rho(\tau) \sim 5\tau^{x^*/2 - 3/2}$ and $\sigma(\tau) \sim -(5/3)x^*\tau^{x^*/2 - 3/2}$. Since $x^* <3$, both $\rho \to 0$ and $\sigma \to 0$   as $\tau \to -\infty$. Thus, each orbit of interest must emerge from the vicinity $P1=(0,0)$ along its unstable manifold.

One can see in Fig.~\ref{portrait} that $U1$, the unstable manifold
of  $P1$, asymptotically tends to
a straight line with slope $-5/9$ corresponding to the Kolmogorov scaling
$F \sim \eta^{-5/3}$. Separatrix $U2$, the unstable manifold
of  $P2$ asymptotically tends to
a straight line with slope $2/3$ corresponding to the thermodynamic scaling
$F \sim \eta^{2}$. Physically relevant solutions correspond to the orbits bound by separatrices $U1$ and $ U2$, and the heteroclinic orbit $H$.

\begin{figure}[h]
 \center{ \includegraphics[width=12cm]{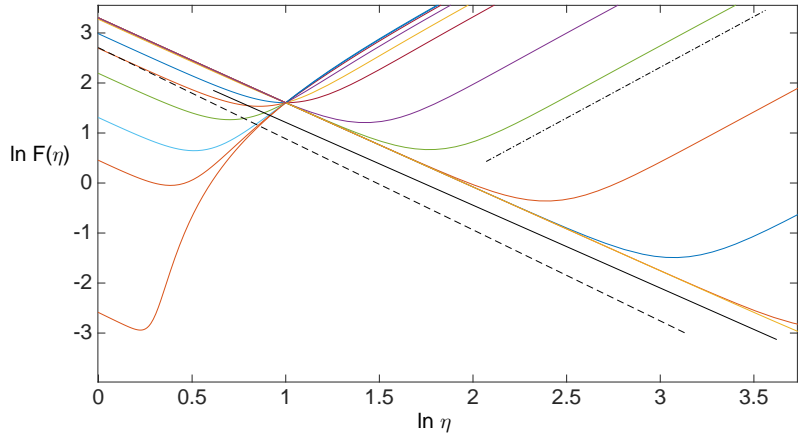} }
\caption{Solutions for $F(\eta)$ corresponding to different degrees of the energy flux reflection. Solid line has slope -5/3. Dashed line has slope $-x^*$. Dash-dotted line has slope 2.}
\label{orbits}
\end{figure}

A typical orbit starts near $P1$, which corresponds to $F \sim \eta^{-x^*}$ at small $\eta$. Then it approaches $U1$ at some intermediate range of $\eta$, which corresponds to Kolmogorov's  $F \sim \eta^{-5/3}$, and then it asymptotes to thermodynamic $F \sim \eta^{2}$ at large  $\eta$;
see Fig.~\ref{orbits}.

To fix a particular solution, one has to specify its behaviour at
large $\eta$. The relevant quantity which can help us to make a choice is the energy flux, which for the model (\ref{invis}) is
\begin{equation}\label{invisF}
\epsilon = - \frac{1}{8}k^{11/2}E^{1/2}\frac{\partial}{\partial k}(E/k^2).
\end{equation}
For the pure Kolmogorov scaling the flux is a positive $k-$independent constant, for the pure thermodynamic scaling it is zero.
Let us fix a vertical line at some large $\rho$ on the $(\rho,\sigma)$-plane, and let us parametrise the orbits by the points of their intersections with this line. Then the lowest lying orbits will be close to the Kolmogorov line, i.e. they will correspond to a constant positive
flux $\epsilon$. The highest lying trajectories will be closest to the thermodynamic line and will have  $\epsilon$ close to zero.
The flux on the orbits lying in between will be monotonically decreasing as we move up our vertical line from the maximum value achieved on
orbit $U1$ to the  zero (asymptotically for $\rho \to \infty$) value achieved on   orbit $U2$.

Physically, the different solutions correspond to different degrees of the energy flux reflection at the large cut-off (or cross-over) wavenumbers.
There is no such cut-off for the classical Navier-Stokes turbulence, and the relevant solution is given by orbit $U1$. This solution does not have a thermalised part. According to this solution, $F(\eta) \sim \eta ^{-x^*}$ for
$\eta \ll 1$ and $F(\eta) \sim \eta ^{-5/3}$ for $\eta \gg 1$. Therefore spectrum $E(k,t)$ has
scaling $\sim k^{-x^*}$ at smaller $k$ and Kolmogorov's $\sim k^{-5/3}$ at the larger $k$, and the point of transition between these two scalings, $k_{tr}$, moves toward the lower $k$ end, $k_{tr} \sim (t-t_*)^b, \, b<0$. Hence the reflected wave scenario
at the final stage of the Kolmogorov spectrum formation at $t_* <t < \infty$.

The extreme case of the complete flux reflection occurs, e.g., in numerical simulations of inviscid (Euler) equations in Fourier space with wavenumber truncation at some $k_{max}$. Formally this corresponds to
orbit $U2$. However, this limit is not so well-posed as orbit $U2$ goes directly to fixed point $P2$, from which it can never leave to move to
fixed point $P1$ and thereby meet the boundary conditions at $\eta \to 0$. This  means that there is no exact self-similar solution that would describe the reflection wave in the case of the complete flux reflection, even though it is perfectly fine to describe cases with strong incomplete reflections.

Incomplete flux reflection occurs, e.g., in numerical simulations of  the fluid equations in Fourier space with some incomplete energy dissipation  near $k_{max}$. This dissipation may be intentional, e.g. via adding a hyper-viscosity term, or simply due to possible dissipative effects related to a particular discretisation algorithm.
As we see in Fig.~\ref{orbits}, stronger flux reflection makes stronger thermalised spectrum and leads to shrinking of the intermediate range exhibiting Kolmogorov's scaling. For very strong reflection   the $\eta^{-x^*}$ range transitions to the thermalised spectrum without any
Kolmogorov range presence in between.

It is interesting that transition to the thermalised range are characterised by presence a range with spectral slopes greater that the thermal value $2$. This has an appearance of a depletion on the spectrum, which is especially pronounced in the case of strong reflections; see
 Fig.~\ref{orbits}. A similar effect was observed in the numerical simulations of the Fourier-truncated Euler equation in~\cite{Cichowlas:2005p1852}. They called such a spectrum depletion  a ``secondary dissipation" attributing its presence to a nonlocal interaction with the thermalised part, the latter arguably giving rise to an effective viscosity effect.
It was further argued that such a feature is impossible within the Leith model as the interactions are very local in $k$ in this case. An indication in favour of this view was the fact that the stationary ``warm cascade" solution
(\ref{01})
does not have such a spectrum depletion.
However, as we can see now, the  depletion does indeed arise within the Leith model when the time-dependent rather than stationary solutions are considered.


\section{Decay of the Kolmogorov spectrum}

Obviously, the reflected-wave self-similar solution will only be physically relevant for a finite time $t=t_* + t_{max}$, namely until  the crossover wavenumber between the $x^*-$range and the $5/3-$range reaches the  scales $\sim k_0$ of the initial spectrum. Just as $t_*$, the value of $t_{max}$ is independent of the viscosity. In fact both of these times are of the order of the turnover
time of the initial eddies, $t_* \sim t_{max} \sim 1/\sqrt{ k_0^3 E(k_0) }$.
At $t=t_* + t_{max}$ one can say that the Kolmogorov spectrum is fully formed: it will be stationary at all later time if there is a permanent forcing at $k_0$.

If there is no forcing in the system, the Kolmogorov spectrum at $t \gg t_* + t_{max}$
will gradually decrease in amplitude as the energy stored near a minimal wavenumber $k_{min}$ (the so-called integral scale) will be gradually bled into larger wavenumbers and dissipated.

The dynamics is still inviscid for $t > t_* + t_{max}$ up to a time $t_{\nu} < \infty$ which we will define later.  The inviscid Leith model
admits a one-parametric family of self-similar solutions of the form
$E(k,t) = t^{3\beta -2} F(k t^\beta)$, where $\beta$ is a parameter~\cite{CHNMG}. The value of the parameter is fixed by the
asymptotics at $k \to 0$. In particular, we can take $E(k,t) \to c_1 k^2$ as $k \to 0$. It is easy to show that the value of the second $k$-derivative of $E(k,t)$ at $k=0$ is conserved by the inviscid Leith model (also by the viscous Leith if $E(0,0)=0$),
constant $c_1$ is time independent.  This  dictates the choice
$\beta =2/5$, Such  behaviour is related to existence of Saffman's invariant, and this is nothing but the scaling suggested by Saffman~\cite{saffman}.
 This is equivalent to taking
$E(k,t) = k^2f(\xi)$ where $\xi = k^{-11/2}t^{-11/5}$ and function $f(\xi)$ satisfies
\begin{eqnarray}\label{E17}
&& \frac{1}{12}(f^{3/2}(\xi))_{\xi\xi} = \frac{11}{5}C^{5/11}\xi^{-6/11}f_{\xi}(\xi),\\
&&  f(0) = 0,\quad f(\infty) = c_1 \label{E18}
\end{eqnarray}
with $C = (2/11)^{-16/11}$. Formally, our self-similar solution $E(k,t)$ behaves as the thermodynamical spectrum for lower wavenumbers.
Note that the problem~(\ref{E17}),~(\ref{E18}) appears in the context of the large time asymptotic of solutions
of the inviscid Leith model after $t_*$, see~\cite{GNMS}.

Interestingly, even though we consider here a solution to the inviscid Leith equation, the total energy decays,
$\int E(k,t) \, dk \sim t^{-6/5}$. This is because of a finite energy flux
to infinite $k$.

In fact, at $t \to \infty$ this process can also be described by a self-similar solution, in this case $E=t^{-1/2} F (t^{1/2} k)$, which is in fact the form inherited from the linear heat equation.
Demonstration of the fact that this is the only possible form of a time-dependent self-similar solution of the
viscous Leith model~(\ref{I1}), as well as the equation for $F(\eta)$, can be found in~\cite{CHNMG}.
According to this solution there is a Kolmogorov scaling range whose minimum and maximum wavenumbers (the integral
and the dissipative scales respectively)
decrease as $\sim t^{-1/2}$, and the total energy decreases as $\sim t^{-1}$.
These are  precisely the laws suggested for the decaying isotropic turbulence
by Lin in 1948~\cite{lin}. Later, alternative laws were suggested, notably by
Saffman, who used conservation of his invariant to derive the decay of the total energy
$\sim t^{-6/5}$. Saffman suggested that the low-$k$ part of the spectrum
scales as $E\sim k^2$, and the Leith model solution predicts the same. The difference in the energy
decay law is explained by the fact that in the Leith model the $k^2$ part has a time-dependent prefactor
in the Leith model solution, whereas it is time-independent in Safmann's model.

\section{Conclusions}

In this paper we considered non-stationary solutions the Leith model of turbulence corresponding to the time $t>t_*$, where $t_*$ is the time
at which the spectral front reaches $k=\infty$ and the first self-similar stage of evolution ends. We found that the spectra at $t>t_*$
is described by self-similar solutions which do not fit into the existing
classification into the first and the second kind of Zeldovich and Raizer
and, therefore, named in the present paper self-similar solutions of the third kind. The latter is defined as a solution whose self-similarity indices can be fixed by neither a conservation law nor by solving an eigenvalue problem, but are determined by an imposed asymptotics at one of the
ends of the similarity interval.

We have obtained a one-parametric family of self-similar solutions corresponding to various strengths of the flux dissipation near a maximal wavenumber. These solutions are generally characterised by
three different power laws having exponent $x^*$ at small $\eta$,
Kolmogorov $-5/3$ at the intermediate  $\eta$ and
thermal $2$ at large $\eta$. There is also a ``secondary dissipation" spectrum depletion between the Kolmogorov and the thermalised ranges,
which was previously found by DNS in~\cite{Cichowlas:2005p1852}.

The most physically important solution in this family, is the one
without a thermalised part. It corresponds to Navier-Stokes turbulence without wavenumber cut-off, in which the energy flux is fully absorbed by viscosity at large wavenumbers without any backscatter.
In this solution the crossover wavenumber between the $x^*$ and the $5/3$ ranges moves toward lower wavenumbers. This
crossover wavenumber can be viewed as the front of a wave reflected off the dissipative scale. It is invading the low-$k$ region  leaving the Kolmogorov spectrum in its wake. Importantly, even though we solve an inviscid problem, the energy is not conserved in this solution. It is decreasing due to a finite flux of energy through the right boundary at an increasing rate, $\epsilon \sim (t-t_*)^{\frac{5-3x^*}{x^* -3}}$.

The reflected-wave  solution is physically relevant for the time $t$ bounded from above by $t_* + t_{max}$.
At this time  the crossover wavenumber between the $x^*-$range and the $5/3-$range reaches the  scales $\sim k_0$ of the initial spectrum.
Both $t_*$ and $t_{max}$ are independent of the viscosity and  are of the order of the turnover
time of the initial eddies, $t_* \sim t_{max} \sim 1/\sqrt{ k_0^3 E(k_0) }$.
At $t=t_* + t_{max}$ one can say that the Kolmogorov spectrum is fully formed: it will be stationary if there is a permanent forcing at $k_0$.
Otherwise it will gradually decrease in amplitude, with its range moving to  smaller wavenumbers, $k_{min}, k_\nu \sim t^{-1/2}$, and
the total energy decreasing as $\sim t^{-1}$ (respectively, $\epsilon \sim
t^{-2}$).

Summarising, the Leith model predicts the following three self-similar evolution stages for the turbulent spectrum which initially has a finite support in the $k$-space. The first stage $t<t_*$ describes a spectral front propagating to arbitrarily large dissipative wavenumber in a finite time $t_*$. The power law spectrum forming behind the propagating front has an anomalous exponent $x^* >5/3$. The second stage at $t_* <t <t_* + t_{max}$ describes a reflection wave from large to small wavenumbers which brings the Kolmogorov spectrum in its wake.
The third stage $t>t_* + t_{max}$ describes a gradual decay of the Kolmogorov spectrum with the Kolmogorov range moving toward smaller
$k$ as $k_{min}, k_\nu \sim t^{-1/2}$.

It is likely that the three-stage scenario of self-similar
evolution is more robust and general beyond the Leith model--
it should hold e.g. for EDQNM model and even DNS of hydrodynamic turbulence. Demonstration of this could be quite a challenging
task remaining for future research.

\section*{Acknowledgement}

This work was partially supported by Grant EPRC (Engineering and Physics Research Council, UK) Fluctuation-driven phenomena and large deviations.

\end{document}